\documentclass{IEEEtran}

\usepackage[utf8]{inputenc}
\usepackage[T1]{fontenc}
\usepackage{amsmath,amsfonts,amssymb,amsthm}
\usepackage{graphicx}
\usepackage{cite}
\usepackage[caption=false,font=footnotesize]{subfig}
\usepackage{epstopdf}
\usepackage{xcolor}
\usepackage{etoolbox}
\usepackage{booktabs}
\usepackage{multirow}
\usepackage[hidelinks]{hyperref}
\usepackage{siunitx}
\graphicspath{ {./figures/} }

\newcommand\hl[1]{#1}

\hypersetup{pdftitle={A Framework for Robust Assimilation of Potentially Malign Third-Party Data, and its Statistical Meaning}}

\newtoggle{eprint}
\toggletrue{eprint}

\title{\huge A Framework for Robust Assimilation of Potentially Malign Third-Party Data, and its Statistical Meaning}
\author{Matthew A. Wright and Roberto Horowitz
\iftoggle{eprint}{}{
\thanks{The authors are with the Department of Mechanical Engineering,
        the California Partners
        for Advanced Transportation Technology (PATH) Program, and
        Berkeley DeepDrive (BDD), University of California, Berkeley,
		CA, 94720, USA (email: \{mwright,horowitz\}@berkeley.edu)}}
        }
        
\begin{document}
\maketitle

\begin{abstract}
    This paper presents a model-based method for fusing data from multiple sensors with a hypothesis-test-based component for rejecting potentially faulty or otherwise malign data.
    Our framework is based on an extension of the classic particle filter algorithm for real-time state estimation of uncertain systems with nonlinear dynamics with partial and noisy observations.
    This extension, based on classical statistical theories, utilizes statistical tests against the system's observation model.
    We discuss the application of the two major statistical testing frameworks, Fisherian significance testing and Neyman-Pearsonian hypothesis testing, to the Monte Carlo and sensor fusion settings.
    The Monte Carlo Neyman-Pearson test we develop is useful when one has a reliable model of faulty data, while the Fisher one is applicable when one may not have a model of faults, which may occur when dealing with third-party data, like GNSS data of transportation system users.
    These statistical tests can be combined with a particle filter to obtain a Monte Carlo state estimation scheme that is robust to faulty or outlier data.
    We present a synthetic freeway traffic state estimation problem where the filters are able to reject simulated faulty GNSS measurements. The fault-model-free Fisher filter, while underperforming the Neyman-Pearson one when the latter has an accurate fault model, outperforms it when the assumed fault model is incorrect.
\end{abstract}

\section{Introduction}
Intelligent transportation systems (ITS) have long relied on the use of real-time data to enable reactive and proactive operations and control.
The widespread and growing use of real-time data, however, brings to ITS a problem that affects many domains in information sciences and engineering: these systems and methods can be fragile when their data are incorrect, either due to faults in the sensors or a feeding-in of malicious data by a hostile attacker (``spoofing'').

\hl{ITS researchers have shown that existing real-time control schemes are sensitive to errors in data.
Some recent research even shows that faulty data can lead to actively harmful control.}
These vulnerabilities exist at both the small-scale, individual-vehicle level, and the multi-vehicle, infrastructural coordinative level.
At the smaller scale, for example, \cite{bhatti_hostile_2017} recently demonstrated the capability to drive a ship off-course via spoofed global navigation satellite system (GNSS) signals, evading detection by both the crew and a statistical spoofing detector.
At the broader, infrastructural level, \cite{reilly_creating_2016} showed how common road traffic control systems and algorithms (e.g., ramp meters and the programs that control the metering rate in response to observed traffic volumes) can be manipulated into causing complex and costly congestion patterns by taking control of their input data.

In this paper, we focus more on the larger-scale end of this spectrum.
Types of ITS applications at this scale include the above-mentioned road traffic control systems, as well as fleet management and tracking systems in industry.
Public and private management entities have both been quick to adopt the use of data from GNSS due to their ubiquitous availability and -- especially for public bodies that wish to avoid the need for expensive installation and maintenance of sensing infrastructure -- relatively low cost \cite{kurvar15}.
Many authors in the ITS community have investigated the use of vehicle-carried GNSS transponders for real-time road traffic observation and control \cite{work2009trafficmodel,lovisari_densityflow_2016,seo_traffic_2017,Ferrara_estimationchapter_2018}.

The work described in this paper was originally inspired by technical problems we encountered in our prior work in this area.
In \cite{wright_pf_2016,wright_cdc_2017}, we report on our efforts to use anonymized \emph{third-party} data from connected vehicles to estimate the state of traffic on a freeway.
That is, the assimilation of records consisting of times, positions, and speeds from transponders near the freeway, but without certainty of the correctness of the data.
For example, upon manual inspection, several records showed transponders with near-zero speeds in times and spaces we believed were not in congestion (e.g., possibly a stopped car), unrealistically fast movement, or speeds that better matched the congestion patterns on the freeway's opposite direction.
\hl{Using a standard particle filter \cite{gordon1993novel} for state estimation,} when some data are of very low probability, led to divergence of the state estimate from the true state, and in some extreme cases, numerical errors caused by floating-point underflow.
In those situations, we want to be able to reject these data that would not improve our state estimate, in a principled manner.

We also sought to develop a method that could reject these malign measurements without having models for all types of faulty data.
This paper describes two modifications to a familiar estimation algorithm, one applicable to the situation where a model of faults exists; and one where such a model does not exist, and the engineer only has a model for sensors' correct behavior.
These two modifications are based on two different mathematical theories of hypothesis testing.

\hl{
In the broader picture, we argue that robustness to faulty data is essential to ITS schemes that make use of GNSS.
ITS schemes often make use of GNSS position, velocity, and time (PVT) measurements, which are susceptible to many sources of error.
These error sources include multipath propagation, non-line-of-sight tracking, signal blockage, tropospheric and ionospheric conditions, and a multiplicity of navigation filter implementations \cite{jin_gnss_2013}.
In other words, the presence of noise or faults in GNSS data for ITS could be considered the norm, rather than the exception, and system robustness to both modeled and unmodeled faults is desirable.
}

% This last point about the need for a fault-detection method without a model of faults draws a parallel between our original motivation and the security context, where experts describe a constant ``arms race'' between attackers and defenders, which motivates the need to be generally robust, including to unknown types of attacks \cite{bhatti_hostile_2017}.
% Attacks to urban GNSS that could stymy ITS are on the horizon: one project funded by the European GNSS Agency towards monitoring interference threats against the GNSS network recently reported at a meeting of the U.S. government's Space-Based Positioning, Navigation, and Timing Advisory Board \cite{dumville_strike3_gnss} of an increase in quantity and technological sophistication of GNSS ``jammer'' threats over the past few years.

The rest of this paper is organized as follows.
Section \ref{sec:filtering} introduces the framework of the \emph{filtering} problem that forms the base for many studies of real-time transportation system estimation \cite{seo_traffic_2017,Ferrara_estimationchapter_2018}, and reviews the popular \emph{particle filter} algorithm that forms the base of our robustified estimation procedure.
% This robustification is based on the familiar mathematical framework of hypothesis testing against incoming measurements.
Section \ref{sec:ht_intro} briefly reviews the theoretical and historical background of hypothesis testing (which forms the core of our robustification), and introduces the two most common frameworks: \emph{Fisherian} and \emph{Neyman-Pearsonian}.
Section \ref{sec:ht_details} goes into the mathematical details of the two frameworks, and describes modifications necessary to apply them to a Monte Carlo scheme like the particle filter.
% The Neyman-Pearson framework works well for the situation where we have a model of fault behavior, and the Fisherian one when we do not.
Section \ref{sec:htpf} merges the standard particle filter with our testing frameworks developed in Section \ref{sec:ht_details}.
Section \ref{sec:case_study} recalls our motivating problem of freeway traffic state estimation using third-party data, and presents some simulation results of the two testing-robustified particle filters on this difficult nonlinear estimation problem.
Section \ref{sec:conclusion} concludes with some discussion on what we feel is this method's interesting fusion of data and model.

\section{Background of the Filtering Problem}
\label{sec:filtering}
\subsection{State Estimation of Dynamic Systems}
We use notation common to nonlinear discrete-time stochastic dynamic systems.
Suppose we have some stateful system whose state evolves in time.
Let $x_k \in R^N$ denote the state vector of the system at time $k$.
The system state is not fully observed; instead what is observed at time $k$ is a measurement vector $y_k \in R^{M_k}$ (the dimensionality having a subscript $k$ implies we may obtain varying numbers of measurements at different times $k$).
The state and observation vectors' temporal behavior are governed by stochastic update and output equations,
\begin{align}
\begin{split}
    x_k &= \mathcal{F}_\theta \left(x_{k-1} \right) \\
    y_k &= \mathcal{G}_\theta \left(x_k \right) \label{eq:fAndgDeterm}
\end{split}
\end{align}
with $\theta$ a parameter vector describing the randomness or process/measurement noise of $\mathcal{F}$ and $\mathcal{G}$.
An equivalent probabilistic notation may rewrite \eqref{eq:fAndgDeterm} as
\begin{subequations}
\label{eq:xAndYAsRVs}
  \begin{align}
    X_k | \left(X_{k-1} = x_{k-1} \right) 
	    &\sim f_\theta\left(x_k | x_{k-1}, \right) \label{eq:fGeneral} \\
    Y_k | \left(X_k = x_k \right) 
	    &\sim g_\theta\left(y_k | x_k, \right) \label{eq:gGeneral}
  \end{align}
\end{subequations}
where, following conventions of probability, a capital letter (e.g., $X_k$) denotes a random vector and a lower-case letter (e.g., $x_k$) denotes the value of a particular realization.
The functions on the RHS's of \eqref{eq:xAndYAsRVs} are probability density functions (PDFs).
More precisely, $f_\theta(\cdot)$ and $g_\theta(\cdot)$ are the PDFs of the conditional distributions for the random variables $X_k$ given $X_{k-1}$ and $Y_k$ given $X_k$, respectively.
%  \cite[Ch. 6]{keener2010theoretical}.

% For the remainder of this paper, we will omit the parameter $\theta$ when our equations begin to get cluttered (i.e., when we are dealing with many PDFs in one equation), but the randomness of the system evolution and observations remain.

The model-based \emph{filtering problem}, a classic problem in stochastic systems, is the problem of estimating the unknown system state $X_k (\forall k)$ from the known observation vectors $Y_k$ \cite{doucet2011tutorial}.
This is often done iteratively forward in time, repeating a two-step process at each successive time $k$.

The first step is called the \textbf{prediction step}.
Assuming that we have an estimate of the PDF of the random variable $X_{k-1} | Y_{0:k-1}$ from the previous timestep, where $Y_{0:k-1}$ is shorthand for the set $\{Y_0, Y_1, \dots, Y_{k-2}, Y_{k-1} \}$, we can use \eqref{eq:fGeneral} to obtain
\begin{equation}
    p_\theta(x_k | y_{0:k-1}) = \int f_\theta(x_k | x_{k-1}) p_\theta(x_{k-1} | y_{0:k-1}) dx_{k-1} \label{eq:theoPredict}.
\end{equation}
The second step is called the \textbf{filtering step} or \textbf{update step}.
Here, we use the obtained measurements $y_k$ and \eqref{eq:gGeneral} to compute \vspace{-5pt}
\begin{equation}
    p_\theta(x_k | y_{0:k}) = \frac{p_\theta(x_k | y_{0:k-1}) g_\theta(y_k | x_k) }{p_\theta(y_k | y_{0:k-1})} \label{eq:theoFilter}
\end{equation}
where
\begin{equation}
    p_\theta(y_k | y_{0:k-1}) = \int p_\theta(x_k | y_{0:k-1}) g_\theta(y_k | x_k) dx_k. \label{eq:margLikelihood}
\end{equation}

Note that \eqref{eq:theoFilter} is a particular statement of Bayes' rule, with $p_\theta(x_k | y_{0:k-1}), g_\theta(y_k | x_k)$, $p_\theta(y_k | y_{0:k-1})$, and $p_\theta(x_k | y_{0:k})$ playing the role of the prior, likelihood, marginal likelihood, and posterior PDFs, respectively.
Because of this, the iterative predict-update approach to filtering is sometimes called \emph{recursive Bayesian estimation} \cite{gordon1993novel}.

For some simple classes of systems $f_\theta(\cdot)$, the computations in \eqref{eq:theoPredict}-\eqref{eq:margLikelihood} are computable in closed form (the most well known example being that if both $f_\theta(\cdot)$ and $g_\theta(\cdot)$ are affine in the state $x_k$ with additive white Gaussian noise, all PDFs in the recursion \eqref{eq:theoPredict}-\eqref{eq:margLikelihood} can be computed exactly through simple matrix algebra, and is known as the Kalman Filter \cite{kalman1960filtering}). In more general settings with more complex system and noise behaviors, some numerical approximation is required.

\subsection{Particle Filter}
One popular approximation method when the integrals in \eqref{eq:theoPredict} and \eqref{eq:margLikelihood} are difficult or computationally intractable is the \emph{particle filter} \cite{gordon1993novel,arulampalam2002PFtutorial,doucet2011tutorial}.
A particle filter may be used even when there is no closed-form expression for $f_\theta(\cdot)$ (precluding many classic numerical integration schemes), but the PDF may be sampled repeatedly, such as by running a stochastic simulation many times.

A particle filter is constructed by replacing the PDFs for $X_k$ in the filtering equations \eqref{eq:theoPredict}-\eqref{eq:margLikelihood} with approximate PDFs, which we will denote with a hat (e.g., $\hat{p}_\theta(\cdot)$ for $p_\theta(\cdot)$).
These approximate PDFs are made up of many discrete samples (also called particles) from the continuous PDF.
The particles are generated by repeatedly sampling from $f_\theta(\cdot)$.

In other words, a particle filter can approximate continuous PDFs via discrete probability mass functions (PMFs).
For example, a particle filter approximation of the posterior PDF $p_\theta(x_k | y_{0:k})$ \eqref{eq:theoFilter} may be written
\begin{equation}
    p_\theta(x_k | y_{0:k}) \approx \hat{p}_\theta(x_k | y_{0:k}) = \sum_{p=1}^P p_\theta(x_k^p | y_{0:k}) \delta_{x_k^p}(x_k) \label{eq:pf_approx}
\end{equation}
where $p \in \{1, \dots, P\}$ denotes individual particles, or atoms of the discrete PMF, and $\delta_{x_k^p}(x_k)$ denotes a Dirac delta that places a unit mass on the point $x_k^p$ (we use the subscript as a notational shorthand for the usual notation, $\delta_{x_k^p}(x_k) \triangleq \delta(x_k - x_k^p)$, where $x_k$ denotes the argument to the ``function'' $\delta(\cdot)$ and $x_k^p$ is the offset that moves the unit mass from $x_k = 0$).

Reviewing the two items in the summand of \eqref{eq:pf_approx}, we see that individual particles have an atom of probability mass placed in the state space of the system, $x_k^p$ (where the superscript $p$ denotes the $p$th particle), and an associated probability $p_\theta(x_k^p | y_{0:k})$.
Summing up these particles results in a PMF with $P$ discrete points, each with an associated probability.

Much like in the theoretical, closed-form version of recursive filtering \eqref{eq:theoPredict}-\eqref{eq:margLikelihood}, the particle filter proceeds in an iterative predict-then-update manner.
As before, to estimate the system state at timestep $k$, we assume that we start with an approximate PDF from the previous timestep, $\hat{p}_\theta(x_{k-1} | y_{0:k-1})$ (note the hat indicating it is an approximation).
This approximation has $P$ individual particles.
We can obtain a particle filter estimate of the prior PDF, $\hat{p}_\theta(x_k | y_{0:k-1})$, by plugging each particle's state value $x_{k-1}^p$ into the stochastic system equation $\mathcal{F}_\theta(x_{k-1})$ \eqref{eq:fAndgDeterm} \cite{gordon1993novel}, \vspace{-5pt}
\begin{equation*}
    x_k^p = \mathcal{F}_\theta(x_{k-1}^p)
\end{equation*}
where the randomness of $\mathcal{F}_\theta(\cdot)$ means that
\begin{equation*}
    \mathcal{F}_\theta(x_{k-1}^p) \sim f_\theta( x_k | x_{k-1}^p).
\end{equation*}
Then, a particle filter approximation for the prior PDF is
\begin{align}
    p_\theta(x_k | y_{0:k-1}) &= \int f_\theta(x_k | x_{k-1}) p_\theta(x_{k-1} | y_{0:k-1}) dx_{k-1} \nonumber \\
    &\approx \sum_{p=1}^P p_\theta(x_{k-1}^p | y_{0:k-1}) \delta_{\mathcal{F}_\theta(x_{k-1}^p)}(x_k) \nonumber \\
    &= \sum_{p=1}^P p_\theta(x_{k}^p | y_{0:k-1}) \delta_{x_k^p}(x_k) \label{eq:pf_predict}\\
    &= \hat{p}_\theta(x_k | y_{0:k-1}) \nonumber
\end{align}
and the particle filter approximation for the posterior PDF is found by plugging \eqref{eq:pf_predict} into \eqref{eq:theoFilter},
\begin{align}
    p_\theta(x_k | y_{0:k}) &= \frac{p_\theta(x_k | y_{0:k-1}) g_\theta(y_k | x_k) }{p_\theta(y_k | y_{0:k-1})} \nonumber \displaybreak[0]\\
    &\approx \frac{\hat{p}_\theta(x_k | y_{0:k-1}) g_\theta(y_k | x_k) }{p_\theta(y_k | y_{0:k-1})} \nonumber \displaybreak[0]\\
    &= \frac{\sum_{p=1}^P p_\theta(x_{k}^p | y_{0:k-1}) \delta_{x_k^p}(x_k) g_\theta(y_k | x_k^p)}
        {p_\theta(y_k | y_{0:k-1})} \nonumber \displaybreak[0]\\
    &= \frac{\sum_{p=1}^P p_\theta(x_{k}^p | y_{0:k}) \delta_{x_k^p}(x_k)}
    {p_\theta(y_k | y_{0:k-1})} \label{eq:pf_filter} \displaybreak[0]\\
    &= \hat{p}_\theta(x_k | y_{0:k}). \nonumber
\end{align}
This posterior approximate PDF is thus made up of the same collection of Dirac deltas as the prior approximate PDF, $\hat{p}_\theta(x_k | y_{0:k-1})$, but with updated weights to reflect each point's posterior probability, after assimilating the measurement $y_k$ through the likelihood.

As has been mentioned, the use of the particle filter avoids having to explicitly calculate difficult integrals.
Of particular relevance is the calculation of the marginal likelihood $p_\theta(y_k | y_{0:k-1})$.
Instead of using \eqref{eq:margLikelihood}, we use \hl{the fact} that in a PMF, the probabilities of all points must sum to one, to normalize the un-normalized probabilities $p_\theta(x_{k}^p | y_{0:k})$ in \eqref{eq:pf_filter},
\begin{equation}
    p_\theta(y_k | y_{0:k-1}) \approx \sum_{p=1}^P p_\theta(x_{k}^p | y_{0:k}). \label{eq:pf_marg_likelihood}
\end{equation}

In implementations of a particle filter, \eqref{eq:pf_filter}-\eqref{eq:pf_marg_likelihood} make up the filtering step that is used in practice.
However, as of yet, we have not brought into consideration the problem of measurement fault detection.
When we introduce the framework for incorporating hypothesis tests for measurement fault detection in Section \ref{sec:htpf}, we will use a different update computation, one that includes an additional hypothesis-testing step.

As an important side note, we have omitted discussion of the particle filter's post-update resampling step because it is not immediately relevant here.
See, e.g., \cite{doucet2011tutorial}, for details.

\section{Statistical Testing: An Introduction}
\label{sec:ht_intro}
Most readers of scientific literature are familiar with hypothesis testing in the form of reports of ``p-values'' and ``statistical significance'' in the context of discussions of, e.g., medical research.
% These types of hypothesis tests generally follow a set procedure.
% First, a null hypothesis $H_0$ is proposed, often a mathematical statement of the converse of the research hypothesis such as ``there is no correlation between these two variables,'' or ``these two sets of data are not significantly different from each other.''
% Then, data is collected and a particular parametric hypothesis test is performed (e.g., the t-test is often used for hypotheses involving the value of the sample mean).
The most popular form of hypothesis test is the so-called ``null hypothesis significance test'' (NHST) \cite{schneider_null_2015}.
\hl{In a NHST, a null hypothesis of, loosely speaking, ``no relation'' or ``no correlation'' is proposed.
Then, a p-value for the data under this null hypothesis is computed, and if it is less than a hard boundary of, e.g., $5\%$, the test is said to have shown ``statistical significance,'' and a specified alternative hypothesis is accepted.}
The NHST is actually a fusion of two distinct theories \cite{schneider_null_2015}: \emph{significance testing}, due to Fisher \cite{fisher1925methods,fisher1935design,fisher1935logic}, and \emph{hypothesis testing}, due to Neyman and Pearson \cite{neyman1928use,neyman1933efficient,neyman1933testing}.

It should be noted that concepts that are rooted in one of the two statistical testing frameworks, but do not make sense in the other, are often discussed alongside each other in the NHST presentation.
For example, the Fisher framework only considers one hypothesis, the null hypothesis.
On the other hand, in the Neyman-Pearson model, multiple hypotheses exist, along with Type I and Type II (also called false positive and false negative, respectively) error rates and statistical power, but p-values are absent (p-values are explicitly defined only in the Fisherian framework) \cite{schneider_null_2015}.

The implications of this dichotomy are more than just philosophical and terminological: for some problems, strict adherence to one theory will lead to a different statistical test than would be derived using the other (see \cite{lehmann_fisher_1993} for more discussion and examples).

\section{Statistical Testing for Measurement Rejection}
\label{sec:ht_details}
\subsection{Notation}
\label{sec:ht_notation}
For this section, where we review classical tests for measurement rejection and introduce new Monte-Carlo-based tests, we will use a somewhat simpler notation.

Suppose that we have data $D = \{d_1, \dots, d_n\}$, which come from a distribution with PDF $p_\theta(D)$.
We use $D$ instead of the classical $X$ for data to avoid confusion with our system state variable.
The PDF has an unknown parameter (or set of parameters) $\theta$.
The testing problem is to evaluate the likelihood of our data for certain values of $\theta$ and make decisions about whether those $\theta$ values should be used or not.

% The notation for our data $D$ will be updated in Section \ref{sec:htpf}, when we combine this section's results with the particle filter.

The remainder of this Section deals with the mathematical details of both the Fisherian and Neyman-Pearsonian theories described above.
We will begin with the Neyman-Pearson framework as its basic elements are likely more familiar to a reader with an applied knowledge of statistics.

\subsection{Neyman-Pearsonian ``hypothesis testing''}
In this framework, in addition to our data $D$ and PDF $p_\theta(D)$, we have two competing hypotheses: $H_0: \theta = \theta_0$ and $H_1: \theta = \theta_1$.
In this case, where both hypotheses fully specify the form of the likelihood $p_\theta(D)$ (since each hypothesis consists of only a single point for $\theta$), a ratio of the two hypotheses' likelihoods might take the form
\begin{equation}
    \Delta(D) = \frac{p_{\theta_1}(D)}{p_{\theta_0}(D)}. \label{eq:likelihoodratio}
\end{equation}
A hypothesis test in this case is often called a ``simple-vs-simple'' hypothesis test (a simple hypothesis is one that fully specifies the model parameters).
For the remainder of this discussion, we will focus on the simple-vs-simple tests.
% It will become apparent in Section \ref{sec:htpf} that this will be sufficient for our needs in this paper.
The formal extension to compound hypotheses is a part of future work.

A well-known result called the Neyman-Pearson Lemma \cite{neyman1933efficient} states that, for a given simple-vs-simple hypothesis testing problem, the optimal test (in that it minimizes the Type II error rate among all tests with a given Type I error rate\footnote{The Type I (``false positive'') error rate, $P(\textnormal{Reject } H_0 | H_0 \textnormal{ true})$, is the mathematical probability that $H_0$ is rejected, conditioned on it being true; and the Type II (``false negative'') error rate, $P(\textnormal{Accept } H_0 | H_0 \textnormal{ false})$, is the probability that $H_0$ is accepted, conditioned on it being false.}) is a \emph{likelihood ratio test}.
A likelihood ratio test is one where the likelihood ratio $\Delta(D)$ is the test statistic of interest.
A likelihood ratio test using the likelihood ratio given in \eqref{eq:likelihoodratio} has the form \vspace{-5pt}
\begin{equation}
    \psi(D) = \begin{cases}
        1 & \textnormal{ if } \Delta(D) < c \\
        0 & \textnormal{ if } \Delta(D) > c
    \end{cases} \label{eq:lrt}
\end{equation}
for some constant $c$.
The test prescribes that, of the two hypotheses, we accept $H_{\psi(D)}$.

The constant $c$ in \eqref{eq:lrt} is chosen to set the likelihood ratio test to have a certain Type I error rate.
This selected Type I error rate is called the test's \emph{significance level} and usually given the symbol $\alpha$.

When the integral is computable, the constant $c$ just mentioned is found by solving for it as a function of $\alpha$ in
\begin{align}
    \begin{split}
    E_{\theta_0} \psi(D) &= \int \psi(D) p_{\theta_0}(D) \; dD \\
    &= P_{\theta_0}(\Delta(D) < c) = \alpha \label{eq:np_level_calc}
    \end{split}
\end{align}
where we use the fact that the expectation of an indicator function (like the likelihood ratio $\psi(D)$ \eqref{eq:lrt}) is equal to the integral of the PDF $p_{\theta_0}$ on the set $\{D: \, \psi(D) = 1\}$.

Of critical importance in \eqref{eq:np_level_calc} is that the PDF of integration for $\psi(D)$ is the likelihood under $H_0$.
This is because the Type I error rate is defined as a rejection of $H_0$ \emph{when $H_0$ is actually true}.
This choice is made because, conventionally, $H_0$ represents the status quo, or a prior belief about $\theta$ before any evidence, and practitioners are interested in tests that have a small error probability when $H_0$ is correct \cite{keener2010theoretical_hyptests}.

Once we have selected our $\alpha$ (and therefore our $c$), we collect the data $D$, evaluate our likelihood ratio \eqref{eq:likelihoodratio}, and select either $H_0$ or $H_1$ depending on the value of $\psi(D)$ \eqref{eq:lrt}.

\subsection{Monte-Carlo Neyman-Pearsonian Likelihood Ratio Tests}
\label{sec:montecarlo_lrt}
We now turn to how we must modify the Neyman-Pearson likelihood ratio test framework for use in our Monte Carlo framework.
Suppose that rather than a known likelihood $p_\theta(D)$, we only have a set of samples $d_i \sim p_\theta(D)$, $i \in \{ 1, \dots, n \}$, as well as their associated likelihoods under the null hypothesis $p_{\theta_0}(d_i)$.
Then, while we cannot solve the integral in \eqref{eq:np_level_calc} in closed form, we can still approximate it via Monte Carlo simulation,
\begin{align}
    \begin{split}
    E_{\theta_0} \psi(D) &= \int \psi(D) p_{\theta_0}(D) \; dD \\
    &\approx \sum_{i=1}^n \psi(d_i) p_{\theta_0}(d_i) \\
    &= \hat{E}_{\theta_0} \psi(D). \label{eq:np_int_montecarlo}
    \end{split}
\end{align}

Designing our test for a specific significance level (i.e., choosing $c$) in this case does not make sense, due to a lack of a closed-form integral equation in which to solve for $c$ as a function of $\alpha$.
Instead, we propose to select either $H_0$ or $H_1$ based on the observed likelihood ratio statistic directly.
The general idea is as follows.
Considering \eqref{eq:np_level_calc} under the classic Neyman-Pearson framework, we would select $c$ so that an $\alpha$ fraction of the probability mass distributed by $\hat{p}_{\theta_0}(D)$ returns values of $D$ s.t. $\psi(D) = 1$ (this is what is shown in the second line of \eqref{eq:np_level_calc}).
Under the empirical approximation \eqref{eq:np_int_montecarlo}, on the other hand, we have a finite amount of points, and can easily compute whether $\Delta(d_i)$ is below or above 1 for every $i$.
The distribution of the probability mass in the empirical distribution is itself defined by our known sample weights $p_{\theta_0}(d_i)$.
\hl{Therefore, we can determine whether at least an $\alpha$ portion of the probability mass under the empirical distribution recommends selecting $H_1$ by simply noting which samples have a higher likelihood under $H_1$ than under $H_0$, and then adding up those samples' $H_0$-probabilities $p_{\theta_0}(d_i)$ and seeing whether this sum is greater or smaller than $\alpha$.}

Mathematically, the empirical Neyman-Pearson likelihood ratio test we propose is
\begin{equation}
    \hat{\psi}(D) = \begin{cases}
        1 & \textnormal{ if } \sum_{i=1}^n 1_{\{\Delta > 1\}}(d_i) p_{\theta_0}(d_i) < \alpha \\
        0 & \textnormal{ otherwise}
    \end{cases}
\end{equation}
where

\begin{equation}
    1_{\{\Delta > 1\}}(d_i) = \begin{cases}
        1 &\textnormal{ if } \Delta(d_i) > 1 \\
        0 &\textnormal{ otherwise}
    \end{cases}
\end{equation}
is an indicator function of whether $d_i$ shows a higher likelihood under $H_1$ than under $H_0$.

\subsection{Fisherian ``significance testing''}
As mentioned above, the Fisherian formulation differs from the Neyman-Pearsonian one in several ways.
One important difference is that it specifies the selection of only a single hypothesis, the null hypothesis $H_0$, which specifies the PDF as $p_{\theta_0}(D)$.
Usually, in a significance test, the null chosen is meant to be more interesting than the common ``no relationship'' hypothesis test, and reflects some \emph{a priori} knowledge.
Observing that the data $D$ do not fit well with the null hypothesis $H_0$ is meant to lead to reconsideration and indication to the practitioner that their prior assumptions used in crafting $H_0$ should be re-evaluated \cite{schneider_null_2015}.

The Fisherian framework calls for the calculation of a test statistic of the data, $T(D)$.
Unlike the simple-vs-simple case in the Neyman-Pearson framework (where we can apply the Neyman-Pearson Lemma), the optimal test statistic is not immediately given.
Instead, its form depends on the particular form of the likelihood $p_{\theta_0}(D)$.

Here, we will assume that the null hypothesis $H_0$ is simple in the Neyman-Pearsonian sense, in that it fully specifies the likelihood: $H_0: \theta = \theta_0$.
Then, we want to compute the tail probability of the observed test statistic $T(D)$ under $H_0$.
This quantity is the p-value.
Whether this tail probability will be a one-sided or two-sided value again depends on the particular form of the PDF $p_\theta(D)$.
For the particular example of a two-sided test, the p-value (assuming $T(D) \in \mathbb{R}$) will be
\begin{align}
    \begin{gathered}
    \textnormal{p-value} = P_{\theta_0}\left(T < -|T(d)|\right) + P_{\theta_0}\left(T > |T(d)|\right) \\
     = \int_{-\infty}^{-|T(d)|} p_{\theta_0}(T(D)) \; dD + \int_{|T(d)|}^{\infty} p_{\theta_0}(T(D)) \; dD \label{eq:twotailed}
    \end{gathered}
\end{align}
where $p_{\theta_0}(T(D))$ is the PDF of the statistic $T$ (also called the statistic's sampling distribution) under $H_0$ and the lower-case formatting of $d$ in $T(d)$ indicates that it is the actually-observed value of the random statistic $T(D)$.

For some common sampling distributions like the univariate Gaussian, Student's $t$, and $\chi^2$ distributions, the solutions to the tail probability integrals in \eqref{eq:twotailed} are available in the familiar statistical testing reference tables, or quickly computed via statistical software.

\subsection{Monte-Carlo Fisherian Significance Tests}
We can move from the theoretical framework of continuous integrals with closed-form solutions \eqref{eq:twotailed} to the Monte Carlo framework using similar arguments as in Section \ref{sec:montecarlo_lrt}.

Considering \eqref{eq:twotailed}, we see the same type of integral we had in \eqref{eq:np_level_calc} (recalling that the test term $\psi(D)$ \eqref{eq:lrt} acted as an indicator function for a one-sided interval, effectively performing the same function as the one-sided limits of integration in \eqref{eq:twotailed}).
Therefore, using similar arguments as before, we can get an approximation of the Fisherian p-value from \eqref{eq:twotailed} (noting the finite integration bound $T(d)$ is replaced with the approximation $\widehat{T}$) as
\begin{equation}
    \widehat{\textnormal{p-value}} = \int_{-\infty}^{-|\widehat{T}|} p_{\theta_0}(T(D)) \; dD + \int_{|\widehat{T}|}^{\infty} p_{\theta_0}(T(D)) \; dD \label{eq:montecarlo_pvalue}
\end{equation}
where
\begin{equation}
    \widehat{T} = \sum_{i=1}^n T(d_i) p_{\theta_0}(d_i). \label{eq:hat_T}
\end{equation}
Like in the Monte Carlo Neyman-Pearson framework, we have again created a weighted-average statistic using our weighted-average approximation of the PDF for the data.

Fisher himself advocated against the use of fixed levels and hard accept/reject boundaries, but instead suggested reporting the p-values directly (for more on this contrast, see, e.g., the discusisons in \cite[Sec. 4]{lehmann_fisher_1993}, \cite[p. 415]{schneider_null_2015}).
However, for our current purposes this is not easily implementable because in our filtering context, we are trying to make a decision as to whether to accept or reject a measurement as we receive it.
Taking a ``soft'' view, and considering a range of state space values based on our current range of belief of whether we should accept or reject $H_0$, while potentially giving us a view of a broader range of possibilities (and separate measurement hypotheses) that we could revisit in light of future data, leads to a blow-up when the number of separate sensors increase \cite{wright_cdc_2017}.
Instead, for expedience, in what follows we adopt the Neyman-Pearson and null hypothesis significance test and select a hard significance level $\alpha$, and reject or accept the measurement based on whether our estimated p-value \eqref{eq:montecarlo_pvalue} is larger or smaller.
A more ``inductive'' approach, closer to the spirit of the original Fisherian significance test, that updates $H_0$ based on repeated tests of measurements from the same sensor, is an avenue for future work.

\section{A Probabilistic Outlier-Rejecting Particle Filter}
\label{sec:htpf}
This section unifies the particle filtering framework reviewed in Section \ref{sec:filtering} and the hypothesis testing methods developed in Section \ref{sec:ht_details}.
As mentioned in Section \ref{sec:ht_notation}, we will unify these ideas in the notation of the state estimation problem.
In this paper, we will not exhaustively define all PDFs of interest in the interest of readability.
See \cite{wright_cdc_2017} for a more lengthy discussion of a precursor to the Fisher-type hypothesis-testing particle filter discussed in the present work.

Recall the filtering or update step in the particle filtering algorithm \eqref{eq:pf_filter}.
In our prior discussion, we considered a measurement likelihood $g_\theta(y_k | x_k)$ \eqref{eq:gGeneral}.
In this notation, we are stating that the random measurement vector $Y_k | X_k$ has a joint distribution across all dimensions.
This makes sense if the measurement noises of the different elements of the measurement vector are correlated or otherwise dependent.

Considering our problem of needing to assimilate data from multiple third-party sensors, though, it makes sense to assume a conditional independence of the measurements.
If we say that at time $k$, we receive measurements from $M$ sensors, with sensor $j$'s measurement being the random variable $Y_k^j$, then by assuming conditional independence of the sensors given $X_k$, we can write
\begin{equation}
    Y_k | (X_k = x_k) \sim g_\theta(y_k | x_k) = \prod_{j=1}^M g^j_\theta(y_k^j | x_k). \label{eq:per_sensor_likelihood}
\end{equation}
This conditional independence assumption is a common assumption in multisensor filtering and sensor fusion (e.g., \cite{chamberland_decentralized_detection_03}, \cite{durrantwhite_multisensor_16}, \cite{mihaylova_freeway_2007})).
Examining \eqref{eq:per_sensor_likelihood}, we notice that we have factored our particle filter likelihood \eqref{eq:pf_filter} into per-sensor PDFs.

Considering a sensor likelihood $g_\theta^j(y_k^j | x_k)$, we can parameterize its nonfault, faulty (for one or more known types of fault, if applicable), spoofed, etc. behavior in $\theta$.
And, if we have models for one of these behaviors, we can accordingly form hypotheses: $H_0: \theta = \theta_0$, $H_1: \theta = \theta_1$, etc.
This is how we bring together the particle-filtering and Monte Carlo fault-detection theories.
Each individual particle, which has a value of the random variable $Y_k^j|X_k$ and a probability (the particle's weighting in the collection of particles), serves as a sample (a $d_i$ in Section \ref{sec:ht_details}'s terminology).
Repurposing these particles as datapoints for the expected behavior of the data under the stated hypotheses lets us reject in real-time measurements that do not match our prior models for data that would come from a correctly-functioning sensor.

Based on the above discussion, the general framework for the robustified particle filter is given below.
% Again, for a more detailed discussion, the reader is referred to \cite{wright_cdc_2017}.
\begin{enumerate}
    \item Perform a prediction step as normal, using \eqref{eq:pf_predict}.
    \item For each sensor $j$ at time $k$, calculate either the likelihood ratio or Fisher test statistic, depending on whether a Neyman-Pearson or Fisher test is used.
    \item For each sensor, determine whether to reject it as faulty using the relevant hypothesis test for the actually-observed measurement and selected $\alpha$.
    \item Perform an update step with the non-rejected measurements using \eqref{eq:pf_filter} (and, if desired, a resampling step).
    \item Advance in time, $k \leftarrow k+1$, return to step 1, repeat.
\end{enumerate}

The type of test (Fisherian or Neyman-Pearsonian) to select for each situation and each sensor is dependent on the problem circumstances.
We believe that it generally makes sense to favor a Neyman-Pearson-type test when one has trustworthy models for all reasonably-expected types of faults, and a Fisherian test when one does not.
Our example case study presented in the next section shows some results for different types of tests and different fault models of varying accuracy.

\section{Example Application}
\label{sec:case_study}
As mentioned in the introduction, our work presented above was motivated by earlier work involving the use of a particle filter and various data sources (some obviously faulty to a human when examined \emph{post facto}) to estimate a freeway's traffic state \cite{wright_pf_2016}.
In this Section, we present a simulation case study based on that work, to demonstrate in particular the hypothesis-testing particle filters proposed in this paper.

\begin{figure*}[ptb]
    \centering
    \subfloat[][True density state (veh/m)]{
      \includegraphics[width=.3\textwidth, trim=0 0 2.5pc 1.5pc,clip]{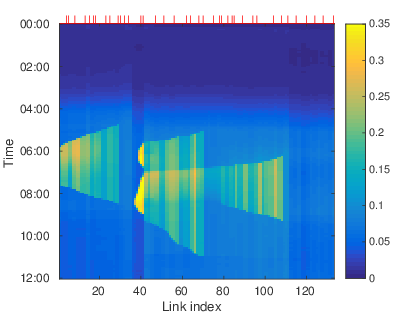}
      \label{fig:c_point_001}
    }
    \subfloat[][Speed measurements (m/s)]{
      \includegraphics[width=.3\textwidth, trim=0 0 2.5pc 1.5pc,clip]{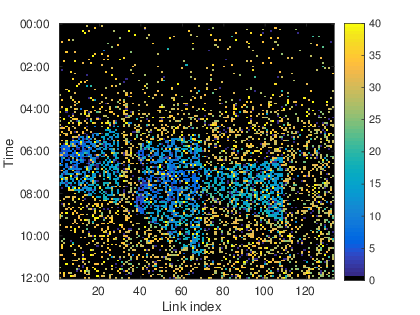}
      \label{fig:c_point_01}
      }
    \subfloat[][Non-faulty speed subset (m/s)]{
      \includegraphics[width=.3\textwidth, trim=0 0 2.5pc 1.5pc,clip]{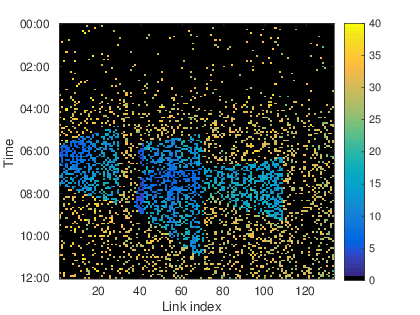}
      \label{fig:c_point_1}
      }
    \caption{Simulated true density state trajectory (a), speed measurements (b), and non-faulty subset of speed measurements (c) used in simulation. Traffic moves to the right, and the time period considered is midnight to noon (as marked on the vertical axis). In (a), the links instrumented with loop detectors that noisily measure density are marked with red ticks. At peak morning demand, bottlenecks near links 30, 70, and 110 lead to traffic jams that propagate upstream (i.e., they extend to the left as time advances), leading to increased density and lower speed. The jams later dissipate as demand falls. \vspace{-12pt}}
    \label{fig:contours}
  \end{figure*}

\subsection{Implementation details}
Our system of study is a 19-mile portion of I-210 West in southern California.
As our system model $f_\theta(\cdot)$, we make use of the macroscopic Cell Transmission Model (CTM) \cite{daganzo94}, which approximates traffic as compressible fluid flows.
This type of model can capture important nonlinear emergent features in traffic flows like traffic jams and congestion waves.

In the CTM, the freeway is discretized into a sequence of finite-volume cells, also called links.
\hl{The state vector $x_k$ is the vector of link densities $\rho_{\ell,k}$, where $\ell$ indexes the links (with $\ell+1$ immediately downstream of $\ell$) and $k$ is the time index.}
% The state update equation for link $\ell$ is
Link $\ell$'s state update equation is
\vspace{-5pt}
\begin{equation}
  \rho_{\ell,k+1} = \rho_{\ell,k} + \frac{1}{L_\ell} (q_{\ell-1,k} - q_{\ell,k} + r_{\ell,k} - s_{\ell,k}), \label{eq:ctm}
\end{equation}
where $L_\ell$ is the length of link $\ell$, $q_{\ell,k}$ denotes the vehicle flow going from link $\ell$ to link $\ell+1$ at time $k$, $r_{\ell,k}$ is the flow entering link $\ell$ from an onramp (if any) at time $k$, and $s_{\ell,k}$ is the flow leaving link $\ell$ to an offramp (if any) at time $k$.

\hl{When there is no onramp or offramp between links $\ell$ and $\ell+1$, the flow from link $\ell$ to link $\ell+1$, $q_{\ell,k}$ in \eqref{eq:ctm}, is given by
\begin{align}
\begin{split}
q_{\ell,k} = \min(& v_{f,\ell} \cdot \rho_{\ell,k} \cdot L_\ell, \, Q_{max,\ell},\\
 &w_{\ell+1} \cdot L_{\ell+1} \cdot (\rho_{J,\ell+1} - \rho_{\ell+1,k})), \label{eq:ctmq}
\end{split}
\end{align}
where $v_{f,\ell}$ is the freeflow speed of link $\ell$, $Q_{max,\ell}$ is the capacity, or maximum possible flow over a time period, of link $\ell$, $w_{\ell+1}$ is the speed at which congestion waves propagate upstream in link $\ell+1$, and $\rho_{J,\ell+1}$ is the jam density, or maximum possible density, of link $\ell+1$.
The third argument in the $\min(\cdot)$ function in \eqref{eq:ctmq} lets the downstream link $\ell+1$ refuse to accept flow from link $\ell$ if $\ell+1$ is too full.}

When there is an onramp and/or an offramp between links $\ell$ and $\ell+1$, we determine the flow $q_{\ell,k}$ according to the junction model of \cite{muralidharan2009lnctm}.
The ramp flows themselves, $s_{\ell,k}$ and $r_{\ell,k}$ in \eqref{eq:ctm}, are random variables.
See \cite{wright_pf_2016} for full implementation details of these last two points.

A common type of first-party sensor for freeway traffic are inductive loop detectors buried in the pavement.
These detectors can noisily measure density.
A third-party source of data are vehicle-carried GNSS devices that report the speed of individual vehicles.
In the CTM, the speed of traffic in link $\ell$ at time $k$ is $v_{\ell,k} = L_l \cdot \rho_{\ell,k} / q_{\ell,k}$.
A high vehicle density leads to congestion, and hence low speeds.
We can use speed measurements to estimate density using this relationship in a Rao-Blackwellized particle filter \cite{doucet2000rbpf}.

To test our fault detection method, we simulated a realization of our freeway model, with randomness introduced by the random onramp, offramp, and upstream boundary flows.
In addition to noisy density measurements from 41 loop detectors, we simulated GNSS speed measurements with a simulated penetration rate of 2\% (i.e., each vehicle had a $2\%$ chance of noisily reporting its speed).
To generate the faulty third-party measurements, we gave each speed measurement a 30\% probability of being faulty.
We used two fault models: a faulty measurement had a 1/3 probability of reporting zero (i.e., a stopped car misreporting its location), and a 2/3 probability of drawing from a Gaussian distribution with mean 30 m/s and standard deviation 10 m/s.
The non-fault model for velocity measurements, $g_{\theta_0}^j(\cdot)$, was Gaussian with a mean of the true link velocity and standard deviation of 20\% of the mean (similar to \cite{work2009trafficmodel}).
Fig. \ref{fig:contours} shows the true state and velocity measurements used.

\subsection{Results}
We tested both the Fisherian and two instances of the Neyman-Pearsonian (each using a different fault hypothesis $H_1$) against this problem.
The results are presented in Tables \ref{tab:np_wrong} through \ref{tab:f}.

\hl{
In the tables, we report both the performance of the statistical tests in rejecting faulty measurements, as well as the resulting estimation error obtained when using the non-rejected measurements.
``Positives'' refer to measurements for which we rejected $H_0$, i.e., sensors that our fault decision statistical test concluded were faulty.
``True'' and ``False'' refer to correct and incorrect decisions, respectively, of whether a measurement is faulty.
``Labeling error'' reports the overall percentage of incorrect decisions (False Negatives and False Positives). }
The estimation error is reported in terms of the mean absolute percentage error (MAPE), the average of $| \hat{\rho}_{\ell,k} - \rho_{\ell,k}| / \rho_{\ell,k}$ for all $\ell$ and $k$, with $\hat{\rho}_{\ell,k}$ the $\ell$th entry of $\hat{\rho}_k = \sum_{p=1}^P x_k^p \cdot p_\theta(x_k^p | y_k)$, i.e., the mean of the posterior particle filter PDF.

The two Neyman-Pearson estimation experiments are shown in Tables \ref{tab:np_wrong} and \ref{tab:np_right}.
As indicated in the table names, Table \ref{tab:np_wrong} shows results for a simulation where the likelihood ratio test had an incorrect faulty measurement likelihood, and Table \ref{tab:np_right} one with the correct faulty measurement likelihood.
The faulty measurement likelihood of Table \ref{tab:np_wrong} was a Gaussian that only placed mass near zero, i.e., it was crafted to select the fake ``stopped car'' vehicles.
\hl{
It indeed rejects the ``stopped car'' vehicles, but, as reflected in the relatively high labeling error rates, it does not reject the purely random measurements.
On the other hand, the Neyman-Pearson fault detector with the correct $H_1$ model, unsurprisingly, performs better, rejecting many of the stopped-car and the purely random measurements.
}
The Fisherian results (Table \ref{tab:f}) show that the estimation accuracy is quite sensitive to the selected $\alpha$.
There are much larger changes in labeling error and MAPE across the scale of $\alpha$ values selected than for either of the Neyman-Pearson results.
This is not too surprising, as not having any alternative hypothesis $H_1$ to compare against, makes the p-value much more sensitive to small variations in the likelihood under $H_0$ than the likelihood ratio would be.

Of particular interest are the columns for $\alpha = 0.001$ and $0.01$ in the Fisher results table (Table \ref{tab:f}).
For these values of $\alpha$, we obtain results that are between the performance of the correct- and incorrect-fault-model Neyman-Pearson filters.
This is an encouraging result, as it confirms our intuition that for a properly tuned $\alpha$, not having a fault model can beat the performance of using an incorrect one.

As mentioned in the caption for the tables, a particle filter estimator that did not see any faulty data \hl{(i.e., the true measurement distribution was $g^j_{\theta_0}(\cdot)$)} obtained a MAPE of 3.43\%.
Unsurprisingly, none of the fault-detecting estimators consistently managed to obtain this level of accuracy, although the Neyman-Pearson fault detector with the correct fault model did come within a standard deviation or two.

\begin{table}[t]
    \caption{Neyman-Pearson Fault Detection/Estimation (Incorrect $H_1$)}
    \label{tab:np_wrong}
    \vspace{-5pt}
    \begin{tabular}{l@{\hskip 2pt}
        S[table-format=4.2(1), tight-spacing=true, zero-decimal-to-integer=true]
        S[table-format=4.2(1), tight-spacing=true, zero-decimal-to-integer=true]
        S[table-format=4.2(1), tight-spacing=true, zero-decimal-to-integer=true]}
      \toprule
      & \multicolumn{1}{c}{$\alpha = 0.001$} & \multicolumn{1}{c}{$\alpha = 0.01$} & \multicolumn{1}{c}{$\alpha = 0.1$} \\ \cmidrule(r){2-4}
      True Positives & 700 (0) & 700 (0) & 700.4 \pm .55 \\
      False Positives & 58.4 \pm 9.50 & 62.8 \pm 5.72 & 76.6 \pm 7.02 \\
      True Negatives & 4535.6 \pm 9.50 & 4531 \pm 5.72 & 1305.6 \pm 7.02 \\
      False Negatives & 1306 (0) & 1306 (0) & 1305.6 \pm 0.55 \\
      Labeling Error (\%) & 20.67 \pm 0.14 & 20.74 \pm 0.09 & 20.94 \pm 0.1 \\ \midrule
      Density MAPE (\%) & 3.80 \pm 0.12 & 3.86 \pm 0.05 & 3.93 \pm 0.11 \\
      \bottomrule
    \end{tabular}
    \vspace{10pt}

    \caption{Neyman-Pearson Fault Detection/Estimation (Correct $H_1$)}
    \label{tab:np_right}
    \vspace{-5pt}
    \begin{tabular}{l@{\hskip 2pt}
        S[table-format=4.2(1), tight-spacing=true, zero-decimal-to-integer=true]
        S[table-format=4.2(1), tight-spacing=true, zero-decimal-to-integer=true]
        S[table-format=4.2(1), tight-spacing=true, zero-decimal-to-integer=true]}
      \toprule
      & \multicolumn{1}{c}{$\alpha = 0.001$} & \multicolumn{1}{c}{$\alpha = 0.01$} & \multicolumn{1}{c}{$\alpha = 0.1$} \\ \cmidrule(r){2-4}
      True Positives & 1415.8 \pm 5.12 & 1433.8 \pm 1.79 & 1450.8 \pm 2.68 \\
      False Positives & 94.8 \pm 8.07 & 106.2 \pm 16.80 & 118.2 \pm 17.61 \\
      True Negatives & 4499.2 \pm 8.07 & 4487.8 \pm 16.80 & 4475.8 \pm 17.61 \\
      False Negatives & 590.2 \pm 5.12 & 572.2 \pm 1.79 & 555.2 \pm 2.68 \\
      Labeling Error (\%) & 10.38 \pm 0.18 & 10.28 \pm 0.28 & 10.20 \pm 0.31 \\ \midrule
      Density MAPE (\%) & 3.51 \pm 0.08 & 3.53 \pm 0.18 & 3.57 \pm 0.17 \\
      \bottomrule
    \end{tabular}
    \vspace{10pt}

    \caption{Fisher Fault Detection/Estimation}
    \label{tab:f}
    \vspace{-5pt}
    \begin{tabular}{l@{\hskip 2pt}
        S[table-format=4.2(1), tight-spacing=true, zero-decimal-to-integer=true]
        S[table-format=4.2(1), tight-spacing=true, zero-decimal-to-integer=true]
        S[table-format=4.2(1), tight-spacing=true, zero-decimal-to-integer=true]}
      \toprule
      & \multicolumn{1}{c}{$\alpha = 0.001$} & \multicolumn{1}{c}{$\alpha = 0.01$} & \multicolumn{1}{c}{$\alpha = 0.1$} \\ \cmidrule(r){2-4}
      True Positives        & 1214.8 \pm 2.49 & 1294.6 \pm 4.62  & 1457 \pm 3.32 \\
      False Positives       & 39 \pm 2.74     & 76.4 \pm 12.10   & 349.4 \pm 29.52 \\
      True Negatives        & 4555 \pm 2.74   & 4517.6 \pm 12.10 & 4244.6 \pm 29.52 \\
      False Negatives       & 791.2 \pm 2.49  & 711.4 \pm 4.62   & 549 \pm 3.32 \\
      Labeling Error (\%)   & 12.58 \pm 0.05  & 11.94 \pm 0.24   & 13.61 \pm 0.49 \\ \midrule
      Density MAPE (\%)     & 3.66 \pm 0.10   & 3.71 \pm 0.18    & 4.22 \pm 0.21 \\
      \bottomrule
    \end{tabular}
    \vspace{3pt}
    
    {\footnotesize ``Positives'' refer to sensors for which we rejected $H_0$, i.e., sensors that our fault detection statistical test concluded were faulty. ``True'' and ``False'' refer to correct and incorrect decisions, respectively, of whether a sensor is faulty. In a simulation with no faulty velocity measurements, a lower-bound density MAPE of \textbf{3.43\%} was achieved.

    All values reported are the mean and standard deviation of five identical simulations with different random seeds.

    MAPE = mean absolute percentage error.}
    \vspace{-15pt}
\end{table}

% Usually, estimating the likelihood of a point under a mixture distribution is done by estimating the probability that the point is drawn from each mixture component, using a method such as the Expectation-Maximization algorithm \cite{EMAlgorithm}, and using those probabilities to create a weighted sum of the likelihoods of the mixture components.
% In this example, we will take a much simpler approach, 

\section{Conclusion}
\label{sec:conclusion}
This article presented a principled fault-detecting particle filter for real-time rejection of potentially faulty measurements.
Our methods are based on the classical Fisherian and Neyman-Pearsonian statistical testing theories, and follows these theories to arrive at different testing methods for when the engineer has a reliable model of faults, and when she does not.

One item of interest is the subtle inversion of what is considered the ``data'' in our hypothesis tests.
Notice that in this paper, the ``data'' that we use to estimate our test statistic are actually the simulated particles, which come from the system model, and we perform our hypothesis test to accept or reject the observations.
An interesting implication of this nuance is how this hypothesis-testing particle filter allows a highly-trusted model to overpower the data, whereas a standard particle filter will always accept every datapoint, even if it is a clear outlier.
We believe that the proposed techniques are closely aligned with the contemporary effort to fuse model-based and data-driven estimation and control techniques in many information sciences: to bring priors obtained from the engineering discipline to the surge of ``big data.''
% Recent efforts in the data-hungry field of reinforcement learning (an area of much interest to ITS due to its promise for autonomous vehicles) have indeed shown that integrating even small amounts of prior knowledge of system dynamics can improve performance \cite{pmlr-v70-chebotar17a}.
These lessons are likely to be useful in many areas to bring GNSS and other ``big data'' to ITS and other built-environment applications.

\section*{Acknowledgements}
This research was supported by the National Science Foundation under grant CPS-1545116 and Berkeley DeepDrive.
M.~A.~W. thanks Adityanand Guntuboyina for raising a question regarding the congruity of \cite{wright_cdc_2017}'s material with the Neyman-Pearson framework that helped direct some of the inquiry reported in the present work.
We also thank our colleague Alex A. Kurzhanskiy as well as the anonymous reviewers for their useful feedback.

\iftoggle{eprint}{
\bibliographystyle{abbrv}
\bibliography{traffic}}{
\bibliographystyle{IEEEtranS}
\bibliography{IEEEabrv,traffic}
}

\iftoggle{eprint}{}{
\begin{IEEEbiography} [{\includegraphics[width=1in,height=1.25in,clip,keepaspectratio]{cropped}}]
{Matthew A. Wright} received the M.S. degree in Mechanical Engineering from the University of California, Berkeley in 2015, and is currently a candidate for the Ph.D. degree in Mechanical Engineering from the same institution.

His research interests include modeling, state estimation, and control for stochastic, complex, and networked systems such as vehicle traffic networks.
\end{IEEEbiography}
\begin{IEEEbiography} [{\includegraphics[width=1in,height=1.25in,clip,keepaspectratio]{Horowitz_Roberto}}]
{Roberto Horowitz} received the Ph.D. degree in Mechanical
Engineering in 1983 from the University of California, Berkeley.
In  1982,  he  joined  the  Department  of  Mechanical Engineering, University of California, Berkeley,
where he is currently the Department Chair and the James Fife
Endowed Chair.

His research interests include the areas of adaptive, learning, nonlinear and optimal control, with applications to microelectromechanical systems (MEMS), computer disk file systems, robotics, mechatronics, and intelligent vehicle and highway systems.

Dr. Horowitz is the recipient of the 2018 ASME Rufus Oldenburger Medal.
\end{IEEEbiography}

}

\end{document}